\documentclass[aps,prb,preprint,superscriptaddress]{revtex4-1}

\usepackage{amsmath} 
\usepackage{graphicx} 
\usepackage{color}
\usepackage{siunitx}
\usepackage[dvipsnames]{xcolor}
\usepackage[normalem]{ulem}

\newcommand{\mr}{\textrm{MR}}

\renewcommand{\vec}[1]{\mathbf{#1}}
\newcommand{\tmin}{T_{\mathrm{min}}}
\newcommand{\tstar}{T^{\ast}}
\newcommand{\tn}{T_{\mathrm{N}}}

\newcommand{\thetap}{\theta^{\prime}}
\newcommand{\vB}{\vec{H}}

\newcommand{\pgroup}{4^{\prime}/m^{\prime} m m^{\prime}}
\DeclareMathOperator*{\spacei}{SI}
\DeclareMathOperator*{\timer}{TR}
\DeclareMathOperator*{\pt}{PT}
\mathchardef\mhyphen="2D

\newcommand{\dphyb}{d\,\mhyphen p}

\definecolor{red}{rgb}{1,0,0}

\begin{document}

\title{
  Itinerant antiferromagnetic BaMn$_2$Pn$_2$'s showing both negative and positive magnetoresistances}

\author{Kim-Khuong Huynh}
\email{huynh.kim.khuong.b4@tohoku.ac.jp}
\affiliation{WPI-Advanced Institute for Materials Research (WPI-AIMR), Tohoku University, 1-1-2 Katahira, Aoba-ku, Sendai 980-8577, Miyagi, Japan}

\author{Takuma Ogasawara}
\affiliation{Department of Physics, Graduate School of Science, Tohoku University, 6-3 Aramaki Aza Aoba, Aoba-ku, Sendai 980-8578, Miyagi, Japan}

\author{Keita Kitahara}
\affiliation{Department of Physics, Graduate School of Science, Tohoku University, 6-3 Aramaki Aza Aoba, Aoba-ku, Sendai 980-8578, Miyagi, Japan}
%


\author{Yoichi Tanabe}
\affiliation{Department of Physics, Graduate School of Science, Tohoku University, 6-3 Aramaki Aza Aoba, Aoba-ku, Sendai 980-8578, Miyagi, Japan}

\author{Stephane Yu Matsushita}
\affiliation{Department of Physics, Graduate School of Science, Tohoku University, 6-3 Aramaki Aza Aoba, Aoba-ku, Sendai 980-8578, Miyagi, Japan}

\author{Time Tahara}
\affiliation{Center for Advanced High Magnetic Field Science, Graduate School of Science, Osaka University, 1-1 Machikaneyama-cho, Toyonaka, Osaka 560-0043, Japan}

\author{Takanori Kida}
\affiliation{Center for Advanced High Magnetic Field Science, Graduate School of Science, Osaka University, 1-1 Machikaneyama-cho, Toyonaka, Osaka 560-0043, Japan}

\author{Masayuki Hagiwara}
\affiliation{Center for Advanced High Magnetic Field Science, Graduate School of Science, Osaka University, 1-1 Machikaneyama-cho, Toyonaka, Osaka 560-0043, Japan}

\author{Denis Ar\v{c}on}
\affiliation{Faculty of Mathematics and Physics, University of Ljubljana, Jadranska c. 19, 1000 Ljubljana, Slovenia}
\affiliation{Jo\v{z}ef Stefan Institute, Jamova c. 39, 1000 Ljubljana, Slovenia}

\author{Katsumi Tanigaki}
\email{tanigaki@sspns.phys.tohoku.ac.jp}
\affiliation{WPI-Advanced Institute for Materials Research (WPI-AIMR), Tohoku University, 1-1-2 Katahira, Aoba-ku, Sendai 980-8577, Miyagi, Japan}
\affiliation{Department of Physics, Graduate School of Science, Tohoku University, 6-3 Aramaki Aza Aoba, Aoba-ku, Sendai 980-8578, Miyagi, Japan}

\begin{abstract}

  We report the discovery of a novel giant magnetoresistance (GMR) phenomenon in a family of BaMn$_{2}$Pn$_{2}$ antiferromagnets (Pn stands for P, As, Sb, and Bi) with a parity-time symmetry.
  The resistivities of these materials are reduced by $60$ times in magnetic fields ($\vec{H}$'s), thus yielding the GMR of about $-\SI{98}{\percent}$.
  The GMR changes systematically along with the Pn elements, hinting that its origin is the spin orbit coupling (SOC) and/or $\dphyb$ orbital hybridization.
  A positive MR component emerging on top of the negative GMR at low temperatures suggests an orbital-sensitive magnetotransport as $\vec{H}$ suppresses the conduction of the electron-like carriers in the $d$-like band but enhances those of hole-like ones in the $\dphyb$ hybridized band.
  The anisotropy of the GMR reveals that the electrical conductivity is extremely sensitive to the minute changes in the direction of the antiferromagnetic moments induced by the parity-time breaking $\vec{H}$, which seems to be associated with a magnetoelectric effect in the dynamic regime of conduction electrons.
  We attribute the observed GMR to the non-trivial low energy band of BMPn's, which is governed by the parity-time symmetry and an magnetic hexadecapole ordering.
\end{abstract}

\date{\today}


\maketitle

\section{Introduction}
\label{sec:intro}

A magnetization ($\vec{M}$) generally responds to a magnetic field ($\vec{H}$), while an electric polarization ($\vec{P}$) to an electric field ($\vec{E}$).
However, there exist cross-coupling phenomena called magnetoelectric (ME) effects in which a magnetic (electric) field can induce an electric polarization (magnetization) \cite{dzyaloshinskii1992,Scott2006}.
Although initially realized only among ferroic materials, the concept of ME couplings is in fact extendable to include antiferroic materials \cite{levitov1985,bychkov1984}.
For instance, in the so-called Edelstein effect, a bulk antiferromagnet with broken time reversal (TR) and space inversion (SI) symmetries is liable to display a spin-polarized current under an external $\vec{E}$ \cite{levitov1985, bychkov1984,edelstein1990}.
The key importance of broken SI and TR symmetries has been highlighted in theory \cite{zelezny2014,zelezny2017}.
A switching of antiferromagnetic (AFM) domains via an electrical current has been experimentally demonstrated in the symmetry-broken itinerant CuMnAs \cite{wadley2016,olejnik2017}.
The physical properties emerging from symmetry-broken itinerant AFM materials have presently become of significant importance.

In this paper we focus on the family of the symmetry-broken BaMn$_{2}$Pn$_{2}$ antiferromagnets (hereafter labeled as BMPn, Pn stands for pnictide elements of P, As, Sb, Bi).
BMPn's are isostructural with BaFe$_2$As$_2$, which is a well-known parent compound for the high-$T_{\mathrm{C}}$ Fe-based superconductors [Figs. \ref{fig:xtals_compare}(a) and (b)].
On the other hand, the G-type AFM order (G-AFM) in BMPn's is distinct from the stripe-type AFM order in BaFe$_2$As$_2$.
In BMPn's, the AFM moments align perpendicular to the two-dimensional MnPn tetrahedral layer, resulting in a magnetic point group of $\pgroup$ that is odd under both SI and TR.
Since the parity-time symmetry is $\pt = \spacei \times \timer$, BMPn's are identified as $\pt$ symmetric materials.
When the vital role of the spin-dependent hybridization between Mn's $d$ and Pn's $p$ orbitals is taken into account, a ferroic  magnetic hexadecapole ordered ground state with the preserved $\pt$ emerges \cite{watanabe2017}.
A good energy matching between the $d$ and $p$ levels places the $\pt$ symmetric state at the top of the valence band \cite{an2009,zhang2016}, making transport measurements the ideal probes for this intertwined magnetic and electronic ground state.
Flexible combination of Mn and pnictogen elements from P to Bi \cite{hoffmann1985} provides an viable way for tuning of spin orbit couplings and $\dphyb$ hybridizations, both of which become larger with the atomic number of the Pn.

 This paper reports our discovery of a new giant magnetoresistance (GMR) phenomenon in the family of small gap semiconducting antiferromagnets BMPn's.
  The origin of this novel GMR effect, which is universal for the whole BMPn family, can be traced back to the ME 
effects in these nearly itinerant materials.
 The magnitude of the magnetoresistance (MR) is greatly enhanced to values as large as $\SI{-98}{\percent}$ 
observed even for moderate fields $\vec{H}$ by changing Pn from As to Bi.
The GMR stems from the ability to manipulate conductivity by breaking the $\pt$,
and hence is the largest when  $\vec{H}$ is perpendicular and the smallest when it is parallel to the direction of zero-field AFM ordered Mn$^{2+}$ moments.
We also show that the $\mr$ exhibits an interesting band-sensitive sign at low $T$'s, in which $\vec{H}$ simultaneously suppresses the conduction of $d$ band carriers but enhances that of the carriers residing in the $d\mhyphen p$ hybridized band.


\section{Results}
\label{sec:results-discussions}
\subsection{BMPn family in the itinerant regime}
All BMPn's are G-type antiferromagnets with very high N\'eel temperatures ($T_{\rm N}$'s), Mn$^{2+{}}$ ion sites being $\SI{625}{\kelvin}$, $\SI{450}{\kelvin}$, and $\SI{387}{\kelvin}$ for BaMn$_2$As$_2$ (BMAs), BaMn$_2$Sb$_2$ (BMSb), and BaMn$_2$Bi$_2$ (BMBi), respectively \cite{yogeshsingh2009,singh2009d,calder2014,saparov2013a,sangeetha2017,ogasawara2018}.
Due to the very small band gaps, BMPn's still reside in the itinerant regime even after AFM ordering sets in.

The G-AFM magnetic order in BMPn's can be understood in terms of a $J_1\mhyphen J_2 \mhyphen J_c$ Heisenberg model justified by the strong Mottness and the localization of the Mn$^{2+{}}$ $3d$ orbitals \cite{johnston2011}.
Here $J_1$ and $J_2$, and $J_c$ are the exchange interactions between the in-plane nearest neighbors and the next nearest neighbors, and the out-of-plane nearest neighbors, respectively.
However, the experimentally measured magnetic moment of each AFM sublattice is only $3.88\,\mu_{\mathrm{B}}$ per Mn$^{2+{}}$ ion \cite{singh2009d,calder2014,sangeetha2017}, which is $\SI{77.6}{\percent}$ of the anticipated fully localized Mn$^{2+{}}$ $S=5/2$ magnetic moment.
  The reduction of the Mn$^{2+{}}$ moments by $\SI{33.4}{\percent}$ is thought to be dispersed from the localized Mn$^{2+{}}$ core via the hybridization with Pn's $p$ orbitals \cite{an2009}.
In this respect, the $d\mhyphen p$ hybridization in the tetrahedral environment is vital for the development of a highly unusual $\pt$ symmetric magnetic hexadecapole ordered state \cite{watanabe2017}.

The intertwined magnetic and electronic ground state in BMPn's is reflected by their non-trivial energetic band structures.
In the low energy region near the top of the valence band, the density-of-states (DOS) for spin-up and spin-down states are almost equivalent despite the G-AFM ordering [Fig.~\ref{fig:xtals_compare}(c)], being evidently unlike the energy-split spin bands found in other ferromagnetic and antiferromagnetic materials \cite{blundell2001magnetism,yang2009c,ma2008,yin2008}.
This unusual spin-dependent DOS in BMPn's, confirmed by {\itshape ab initio} calculations \cite{an2009,zhang2016}, can be understood as a manifestation of the $\pt$ symmetry, under which particles with opposite spins are degenerated at every point in the momentum space even though the Kramer degeneracy is forbidden.
Second, the aforementioned spin-dependent hybridization results in a partially itinerant valence bands containing $\dphyb$ mixing orbital characters, whereas the conduction band remains almost in pure $d$-states.
In both aspects of spin and orbital, the valence band differs remarkably from the conduction band, the latter of which contains mostly spin-down $d$-like states.
The electrically excited charged quasi-particles occupying these fundamental bands are tightly coupled to the multipolar ground state \cite{watanabe2017}.
The magnetic hexadecapole state can be altered by $\vec{H}$ and intriguing magnetotransport properties are expected.

\subsection{Comparison of BaFe$_2$As$_2$ and BMAs magnetoresintaces}
\begin{figure*}
  \centering
  \includegraphics[width=.6\textwidth]{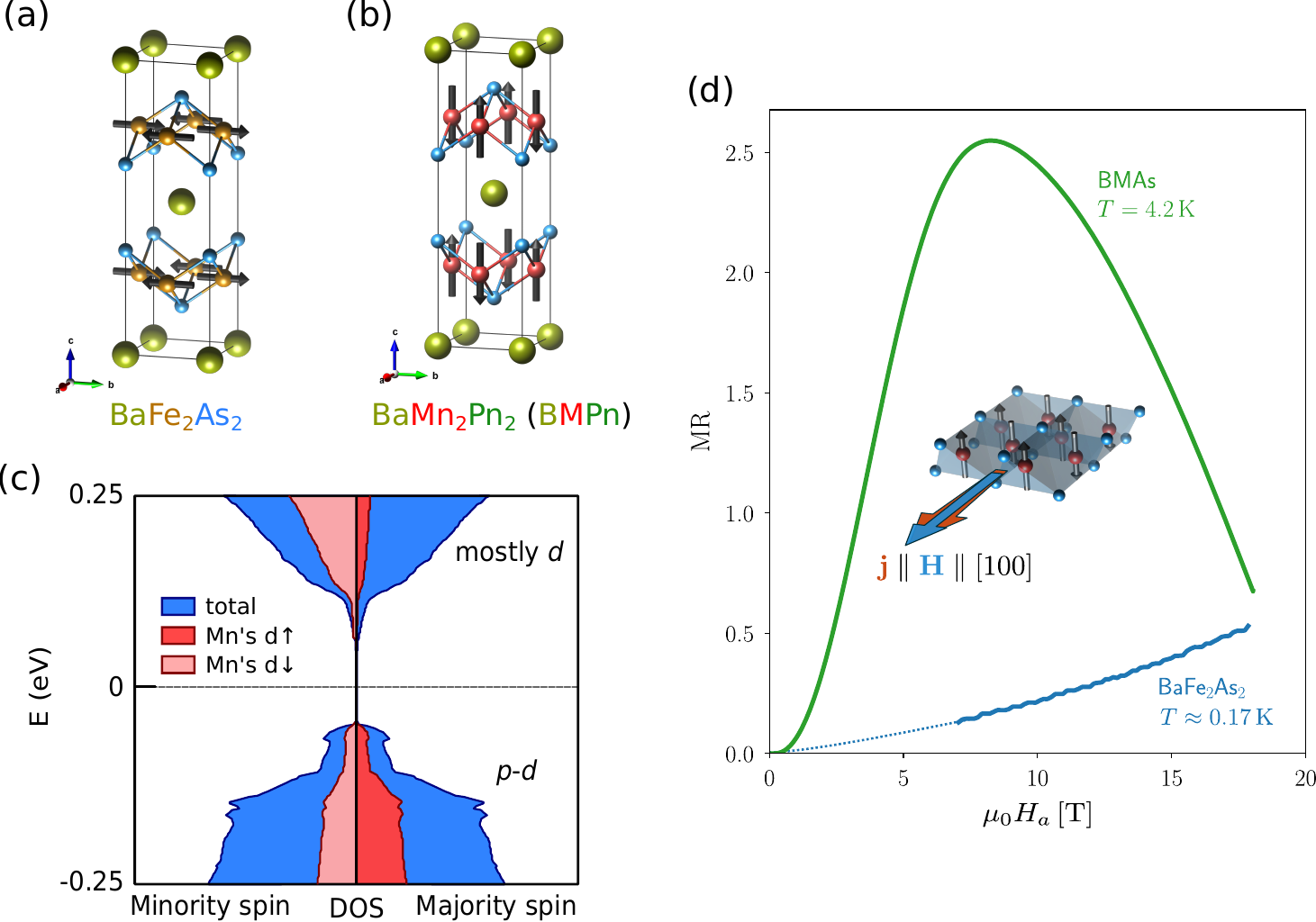}
  \caption{
    (a) and (b) the crystallographic and the magnetic structures of BaFe$_2$As$_2$ and BMPn's.
    Whereas BaFe$_2$As$_2$ and BMPn's adopt the same crystallographic space group $I4/mmm$, the magnetic orders are different (see text). 
    BMPn's show a G-type (checkerboard) AFM order of almost localized Mn$^{2+}$ moments along the $c$-axis.
    (c) Schematic spin-polarised band structure of BMPn's (adapted from \cite{an2009}).
    (d) Longitudinal $\mr$'s of BMAs and BaFe$_2$As$_2$ with $\vec{H}$ aligned parallel to both $\vec{j}$ and the $\hat{a}$ axis as shown in the inset.
    The longitudinal $\mr$ observed in a best quality single crystal of BaFe$_2$As$_2$ at $T = \SI{0.17}{\kelvin}$ (taken from Ref.~\onlinecite{terashima2011}) exhibits the conventional $H^2$-dependencer, which is generally predictable in the framework of the semiclassical transport theory \cite{pippard1989}.
    This is in a sharp contrast to the unconventional bell-shape $\mr$ observed in BaMn$_2$As$_2$ (BMAs).
    The dotted line is a guide to the eye.
  }
  \label{fig:xtals_compare}
\end{figure*}

In Fig.~\ref{fig:xtals_compare}(d) we compare the low-temperature magnetotransport properties between BMAs and  
and its sister compound BaFe$_2$As$_2$. 
Magnetotransport properties of BaFe$_2$As$_2$, which is a compensated multiband semimetal with tiny Dirac cone states \cite{huynh2011,huynh2014}, generally obey the law of semiclassical transport theory.
In particular, magnetoresistance ($\mr$), defined as $\mr = \rho(H)/\rho(0) - 1$ where  $\rho$ is the resistivity, is minimized when $\vec{H}$ is aligned parallel to the current density ($\vec{j}$) and monotonically and gradually increases with $H^2$ \cite{pippard1989}.
In contrast, in the case of BMAs, measurements in the same longitudinal configuration ($\vec{H} \parallel \vec{j} \parallel \hat{a}$) yield an unprecedentedly large $\mr$.
Moreover, the $\mr$ exhibits a 
bell-shaped field dependence: 
MR at low magnetic fields first increases with $H_a$ but then reaches a maximum value of around $250\,\%$ at $\mu_0 H_a \approx \SI{8}{\tesla}$ (here $H_a$ is the component of $\vec{H}$ along the $\hat{a}$-axis, see Sec. Experiments).
On further increase of magnetic field $\mr$ starts to decrease and eventually becomes negative for $\mu_0 H_a \sim \SI{23}{\tesla}$ and then saturates at around $-98\,\%$ under $\mu_0 H_a \gtrsim \SI{35}{\tesla}$ [Fig.~\ref{fig:BMPn_MRs}(b)].
This demonstrates that MR of BMAs is very complex as a function of $\vec{H}$ and shows both positive and negative contributions to the total MR.

Although such shape of $\mr$ in Fig. \ref{fig:xtals_compare}(d) may look similar to those originating from the well-known magnetic breakdown phenomena in metals, we emphasize that the mechanism of magnetic breakdown does not allow negative $\mr$ \cite{pippard1989} and hence it is not applicable to our case.
Furthermore, as shown later in Fig.~\ref{fig:anisotropicMR}, angle-resolved MR measurements of BMPn single crystals under $\vec{H}$ indicate a very unique anisotropy of the MRz that cannot be understood using any available theoretical models so far proposed.

\subsection{ Tuning of the giant magnetoresistive effect via Pnictogen elements} 
\label{sec:tuning-gmr-via}
\begin{figure}
  \includegraphics[width=.35\textwidth]{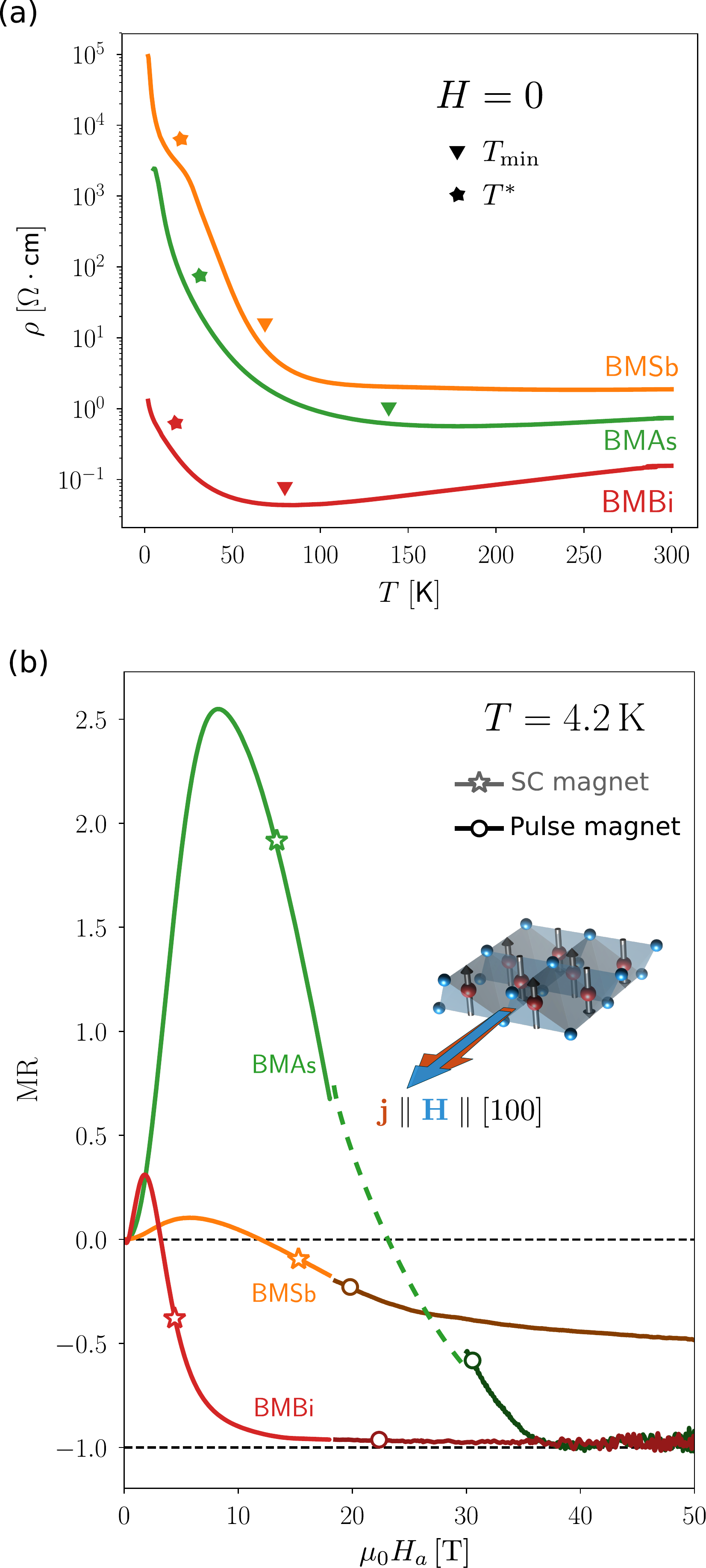}
  \caption{
    (a) Temperatures dependencies of in-plane $\rho$'s in BMAs, BMSb, and BMBi single crystals under zero $H$.
    Each of these three compounds shows a crossover from bad a metal to an insulator at $\tmin$ (marked by a inverse triangle) and an additional kink at $\tstar < \tmin$ (marked by a star).
    (b) GMR's in BMAs, BMSb, and BMBi measured at $T=\SI{4.2}{\kelvin}$ under a longitudinal geometry where both $\vec{H}$ and $\vec{j}$ were aligned along the $\hat{a}$-axis (see the inset), showing a universal positive-negative-saturated shape and with a clear Pn-dependence.
    In each curve, the data collected using a $\SI{18}{\tesla}$ superconducting magnet (lighter color marked with star) were matched with those measured using a $\SI{55}{\tesla}$-pulse magnet (darker color marked with a circle).
    The dashed line is a guide to the eye.
}
  \label{fig:BMPn_MRs}
\end{figure}

BaMn$_2$P$_2$ (BMP) is highly insulating with very high $\rho$ already at room temperature and  almost immeasurable $\rho$ at cryogenic $T$'s. 
Although the other three members of the BMPn family are also highly resistive at low $T$'s [Fig.~\ref{fig:BMPn_MRs}(a)], they have much more tolerable values of $\rho$ that allowed us to study their electrical transport in a wide range of $T$.
The isothermal longitudinal $\mr$ of BMAs, BMSb, and BMBi measured at $T=\SI{4.2}{\kelvin}$ are summarized in Fig.~\ref{fig:BMPn_MRs}(b).
For all three samples, MR showed similar bell-shaped field-dependent MR features as those of BMAs described above.
However, characteristic values of $\mr$, i.e., the field of maximum in the positive MR and the field at which the negative MR saturates, systematically change with the Pn elements from the lighter As element to the heavier Bi one.

Compared to BMAs, BMSb shows an intermediate magnitude of $\mr$ which saturates at around $-50\%$ for $\mu_0 H > \SI{25}{\tesla}$.
The most interesting is the case of BMBi for which whole features of $\mr$ are observable in the smallest $H$-range.
Since the positive $\mr$ component seems to be the smallest for this compound, $\mr$ reaches a saturation value of $-98\,\%$ already for $\mu_0 H\approx\SI{18}{\tesla}$, which corresponds to a 60-times reduction of the resistivity.

We emphasize that the magnitude of negative $\mr$ is very high. 
For example, the field induced metal-to-insulator transition in the manganite colossal magnetoresistance (CMR) systems results in $\mr$ of around $-99\,\%$ under the similar $H$ strength \cite{tokura1994,kuwahara1995}.
The remarkable similarity in the shape of the $\mr$ field-dependence across the BMPn's indicates the manifestation of a unified physical mechanism. 

\subsection{The angular-dependence of the giant magnetoresistive effect}
\label{sec:anis-giant-magn}

\begin{figure*}
  \includegraphics[width=.8\textwidth]{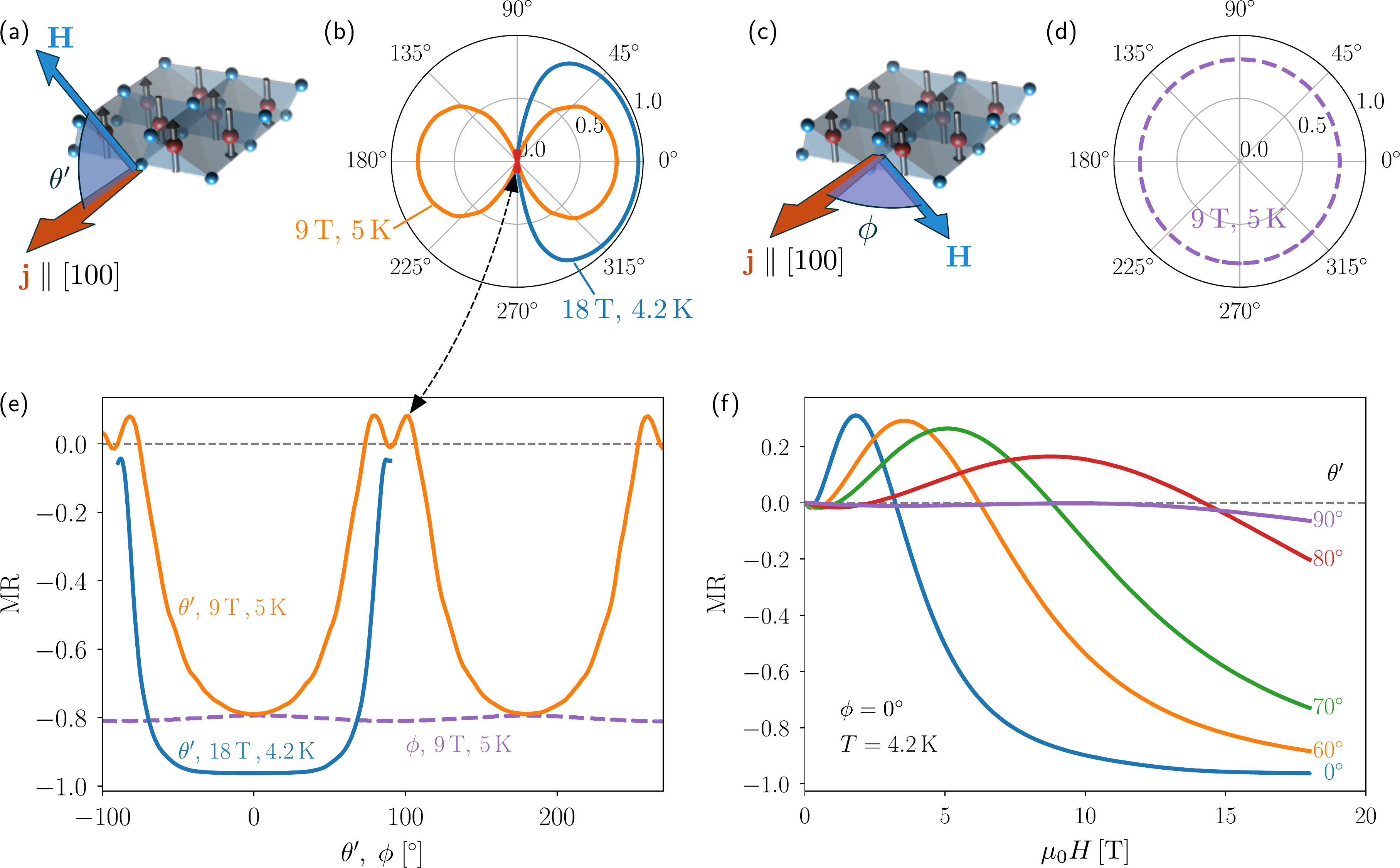}
  \caption{
    Anisotropy of $\mr$ in BMBi.
    (a) and (c) illustrate the out-of-plane and in-plane rotations ($\theta^{\prime}$ and $\phi$) of $\vec{H}$ with respect to the $\hat{c}$-axis.
    (b) shows the dependencies of $\mr$ on $\theta^\prime$ under $\SI{5}{\kelvin},\,\SI{9}{\tesla}$ and $\SI{4.2}{\kelvin},\,\SI{18}{\tesla}$ in polar plots, in which the angular dependence of the radius represents that of the absolute value of $\mr$.
    The $\theta^{\prime}$-rotations indicate very large angular dependencies of $\mr$, which is in a sharp contrast to an almost isotropic $\mr$ observed in the $\phi$-rotation shown in (d).
    (e) The same data as those in (b) and (d) in linear plots.
    The positive part in the $\mr$ at $\SI{5}{\kelvin},\,\SI{9}{\tesla}$ corresponds to the small red segments in (b), as indicated by the dashed arrow.
    Under strong $H$'s, $\mr$ is almost constant as $\thetap$ varies in the range of $\pm\SI{20}{\degree}$ around $\SI{0}{\degree}$ or $\SI{90}{\degree}$, resulted in the unusual clam-shell shape in the polar plot (a).
    On the other hand, the $\phi$ rotation indicates an almost isotropic MR.
    (f) $H$-dependencies of the isothermal $\mr$'s at $T = \SI{4.2}{\kelvin}$ measured at $\phi = \SI{0}{\degree}$ and various $\thetap$'s.
  }
  \label{fig:anisotropicMR}
\end{figure*}

The $H$-dependencies of longitudinal $\mr$ described above clearly show that BMBi is the key member of the BMPn family on which more elaborated experiments should be made.
We therefore proceeded to measure the angular dependence of $\mr$ and $\rho(T)$ on BMBi single crystals for different static $H$'s.
When $\vec{H}$ was rotated in the crystal $ac$ plane, the $\mr(\theta^\prime)$ curves exhibited an unusual clamp-shell shape in the polar plot [Fig.~\ref{fig:anisotropicMR}(b)]. 
When $\vec{H}$ was rotated away from the $ab$ plane (i.e., when $\thetap$ increased), the isothermal $\mr$ curves were elongated towards the high $H$ region and eventually became negligibly small when $\vec{H}$ was perpendicular to the $ab$ plane.

Measuring the in-plane anisotropy of MR by rotating $\vec{H}$ in the $ab$ plane by the angle $\phi$ yields the polar plot of $\mr(\phi)$, which is a nearly perfect circle [Fig.~\ref{fig:anisotropicMR}(d)].
Very small modulations that give maxima in $\mr$ are detected for $\vB \perp \vec{j}$ [Fig.~\ref{fig:anisotropicMR}(e)].
Putting side by side with the strong out-of-plane anisotropy described above, the in-plane isotropy observed here delivers an important message on the symmetry of the unique $\mr$.
The mechanism that induces the large $\mr$ observed in BMPn is sensitive to the angle between $\vB$ and the $\hat{c}$ axis, and it is nearly unaffected by the relative angle between $\vB$ and $\vec{j}$ as long as $\vB$ is parallel to the $ab$ plane.
Measurements of $\mr$ under $\vB$ rotating in the $ac$ plane (not shown) indicate an angular dependence being almost identical to what can be seen in the $\thetap$-resolved experiments, thus corroborating the conclusion on the symmetry of $\mr$.


\subsection{The three temperature regimes in magnetotransport properties of BMPn}
\label{sec:t-regim-magn}
\begin{figure*} 
  \includegraphics[width=0.6\textwidth]{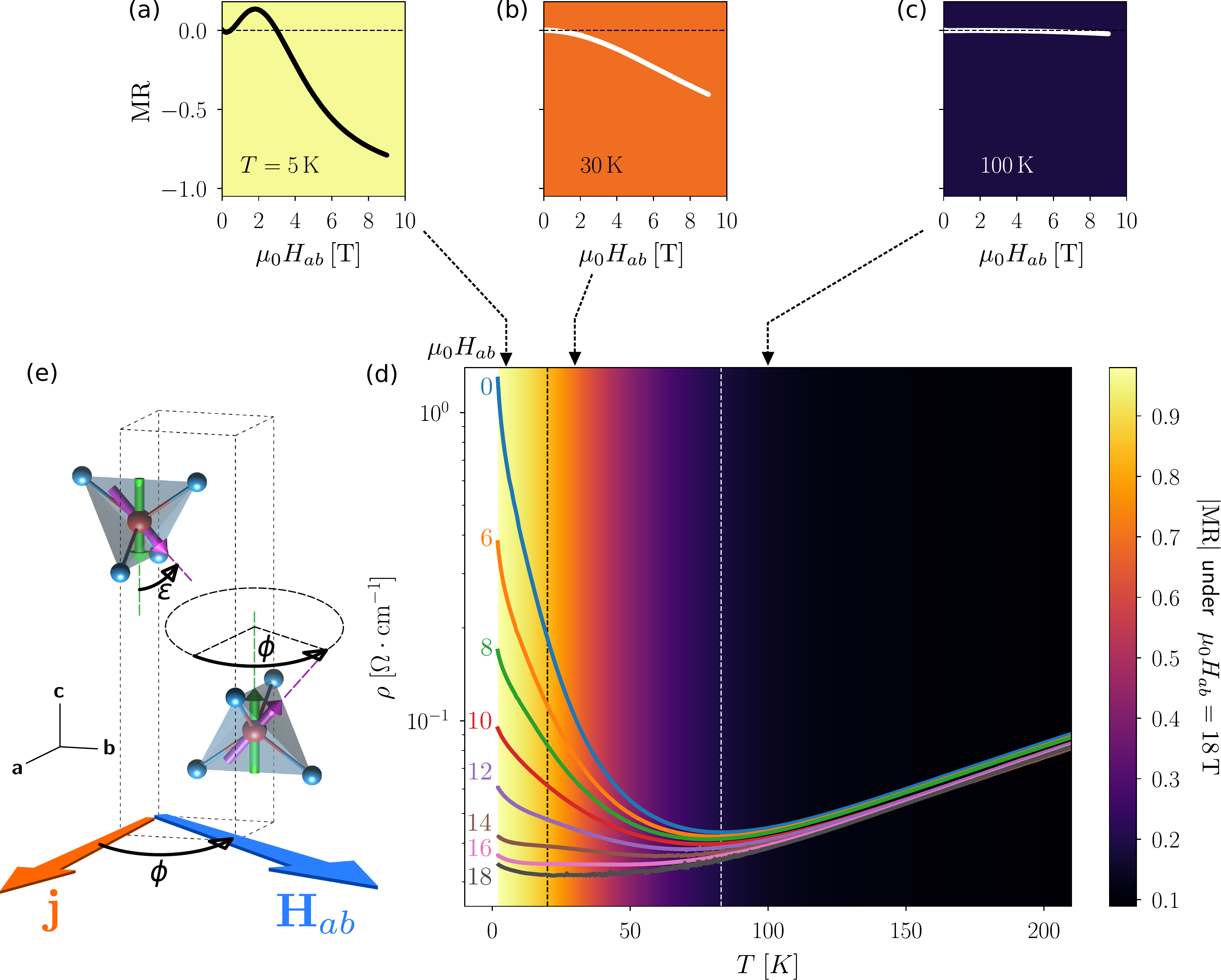}
  \caption{Three $\mr$ regimes in BMBi.
    Figures (a), (b), and (c) show the characteristics $H_{ab}$-dependencies of $\mr$ in the LT, MT, and HT regimes, respectively.
    (d) shows $T$ dependencies of $\rho$ in BMBi measured under various static $H_{ab}$'s.
    The value of $H_{ab}$ for each curve is indicated by the number at the left hand side.
    The background shading of (d) is a mapping of the $T$-dependence of the absolute value of $\mr$ under $\mu_0H_{ab}=\SI{18}{\tesla}$.
    Two dashed lines at $T^{\ast}=\SI{20}{\kelvin}$ and $\tmin=\SI{83}{\kelvin}$ separate three regimes of $\mr$.
    (e) An illustration for the effect of the in-plane $\vec{H}_{ab}$ on the directions of the Mn$^{2+}$ AFM spins.    
 }
  \label{fig:BMB_Tdep}
\end{figure*}

The G-AFM phase transition at high $\tn$'s in BMPn's leads to the opening of a small electronic insulating gap below $\tn$.
This is supported by the first principle calculations \cite{an2009,mcnally2015,zingl2016} and experimentally verified in the angle-resolved photoemission spectroscopy (ARPES) \cite{zhang2016}, and thermodynamic and magnetic measurements \cite{yogeshsingh2009,saparov2013a,sangeetha2017,ogasawara2018}.
Although temperature evolution of the longitudinal resistivity shows anomalies at two temperatures below the G-AFM phase transition temperature, 
no anomalies have been detected so far in a wide range of $T$ below $\tn$ \cite{yogeshsingh2009,saparov2013a,ogasawara2018,sangeetha2017}.

In sharp contrast to these earlier reports, our electrical transport data measured on BMBi as a function of $\vec{H}$ and $T < \tn$ reveal a remarkably rich phase diagram that comprises several regions characterized by different regimes in resistivity.
Under zero $H$, the $T$ dependence of the in-plane $\rho (T)$ at high $T$s still follows a bad metal-like behavior despite the opened band gap in the G-AFM phase [Fig.~\ref{fig:BMB_Tdep}(d)].
$\rho(T)$ reaches a broad minimum at $\tmin \approx \SI{83}{\kelvin}$ and on further cooling it starts to follow a standard Arrhenius law $\rho \propto \exp{(E_{\mathrm{A}}/{2k_{\mathrm{B}}T})}$ with an activation energy $E_{\mathrm{A}}\approx \SI{8}{\milli \electronvolt}$.
However, the activation behavior in  $\rho(T)$ can be traced only in a narrow $T$-window between $\tmin$ and $\tstar \approx \SI{20}{\kelvin}$.
The kink observed at $\tstar$ is comparatively more subtle in the case of BMAs and BMBi, but it is clearly pronounced in the case of BMSb, most probably because BMSb is the most insulating compound among the three. 
The two temperatures $\tmin$ and $\tstar$ virtually divide the G-AFM state into three regions below $\tn$: the high $T$ (HT): $\tmin < T < \tn$, the middle $T$ (MT): $\tstar < T < \tmin$, and the low $T$ (LT) regime: $T < \tstar$.

The BMPn materials exhibit distinct magnetotransport properties in each $T$-region as schematically summarized in Fig.~\ref{fig:BMB_Tdep}(a)-(c) for BMBi as an archetypal member of the BMPn family.
In the HT regime, we observed almost $T$ independent small negative $\mr$'s, which reaches the value of $-10\,\%$ for in-plane $H_{ab}$ as strong as $\SI{9}{\tesla}$ [Fig. 4(c)].
The magnitude of the negative $\mr$ becomes substantially larger with decreasing $T$.
This is clear for various $H_{ab}$'s, where negative $\mr$ become quite pronounced in the MT regime as shown in Fig.~\ref{fig:BMB_Tdep}(b).
Compared to the HT regime, $\mr$ is now an order of magnitude larger.
Finally, the $\mr$ exhibited a complex bell-shape $H_{ab}$-dependence decorated with competing positive and negative components in the LT regime [Fig.~\ref{fig:BMB_Tdep}(a)].
For example, at $T = \SI{5}{\kelvin}$ the positive $\mr$ dominates at lower $H_{ab}$'s, resulting in a maximum of $+30\,\%$ in magnitude at $\mu_0 H_{ab}\approx\SI{2}{\tesla}$.
For higher $H_{ab}$'s, a large and monotonic decrease to the negative side was observed. 
The $\mr$ gradually saturates at $\mu_0 H_{ab}\approx\SI{6}{\tesla}$, and eventually $\rho$ is reduced by $60$ times at $\mu_0H_{ab}$ of $\SI{18}{\tesla}$, pushing the material back into a metallic state [Fig.~\ref{fig:BMB_Tdep}(d)].
The positive maximum in $\mr$ in the LT regime declines with elevating $T$'s and virtually disappears as the negative component of $\mr$ prevails in the MT regime. 
On the other hand, the negative part of $\mr$ under high $H_{ab}$ sustains until very high $T$'s with holding its intensity to be constant.
In the whole $T$-range, the anisotropy of the MR effects is unchanged, i.e., MR's show maximum for $\vec{H} \parallel ab$ and are almost isotropic when $\vec{H}$ is rotated inside the $ab$ plane.
On the other hand,  $\vec{H}$ parallel to the $\hat{c}$ axis minimizes the MR effects.

\subsection{$\mr$ and transverse Hall effects in low-temperature regime}
\label{sec:transv-hall-effects}

\begin{figure} 
  \includegraphics[width=.4\textwidth]{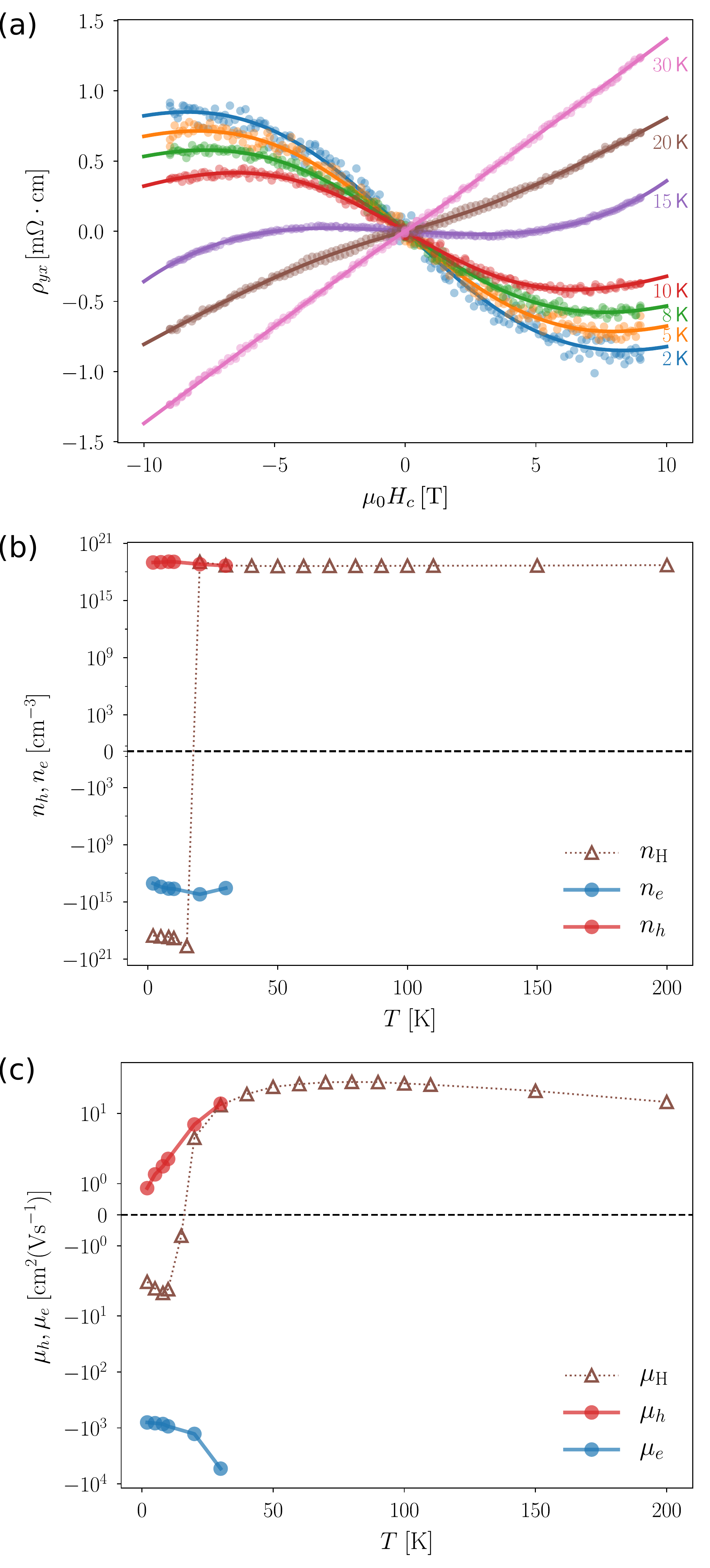} 
  \caption{
    Transverse Hall effects of BMBi single crystal.
    (a) Evolution with $T$ of the Hall resistivity $\rho_{yx}(B)$.
    The lines represent the fittings of a two-carrier-type model to the data (points).    
    Numbers (b) and mobilities (c) of electron-like and hole-like carriers extracted from the two-carrier-type model.
    Here $n_h$ and $\mu_h$, and $n_e$ and $\mu_e$ denote the numbers of and mobility of hole-like and electron-like carrier types, respectively.
    For comparison, the carrier number $n_{\mathrm{H}}$ and mobility $\mu_{\mathrm{H}}$ obtained from the one-carrier fitting at low $H$ region was also included.
  }
  \label{fig:BMB_Hall}
\end{figure}

The emergence of the large positive longitudinal MR component below $T^\ast$ is indeed puzzling.
This becomes even more enigmatic given the experimental fact that other measurements, such as magnetic susceptibility and specific heat, could not detect any anomaly in this temperature range \cite{yogeshsingh2009,saparov2013}.
No anomalies in the $T$-dependence of specific heat suggest that if there is a thermodynamic transition, then its order should be higher than 2.

Interestingly, our measurements of the transverse Hall effects indicate an emergence of electron-like carriers for $T \lesssim T^{\ast}$.
As shown in Fig. \ref{fig:BMB_Hall}(a), in both HT and MT regimes, the $H_c$ dependence of Hall resistivity $\rho_{yx}(H_c)$ is positively linear, indicating the dominant contribution of single hole-like carriers.
As $T$ approaches $T^{\ast} \approx \SI{20}{\kelvin}$, non-linearity develops in $\rho_{yx}(H_c)$ and in the low $H_c$ region a slope bends to the negative side.
At the lowest $T$'s, the shapes of $\rho_{yx}(H_c)$ resemble those observed in multiple carrier systems with a negatively linear segment at low $H_c$, followed by a non-linear bending towards the positive side as $H_c$ increases.
In order to extract both the carrier numbers and the mobilities, we analyzed the transverse Hall data by isotropic semiclassical models, which include either one or two carrier types.
The analyses displayed in Fig.~\ref{fig:BMB_Hall}(b) and (c) indeed confirm the coexistence of both electron- and hole-like carriers for $T \lesssim T^{\ast}$. 
The mobility of electron-like carriers can be estimated to be around $\SI{1e3}{\square\centi\meter\per\volt\per\second}$ at $T = \SI{2}{\kelvin}$, which is much higher than that of the hole-like ones, being around $\SI{1}{\square\centi\meter\per\volt\per\second}$ [Fig.~\ref{fig:BMB_Hall}(c)].
The influence of high mobility electron-like carriers is thus evident even though their number is small in comparison with the pre-existing hole-like ones.


\section{Discussion}
\label{sec:discussions}

The main result of this work -- a discovery of the giant magnetoresistance in BMPn family -- already discloses its main features: (1) the magnitudes of the unique negative $\mr$'s of BMA and BMB in the high magnetic fields states are larger than $\SI{98}{\percent}$, which are the values comparable to those observed in other exotic GMR materials, such as manganites \cite{tokura1994,kuwahara1995} and Weyl metals \cite{son2013,hirschberger2016,xiong2015,lu2016}, (2) GMR exhibits a very unusual angular dependence under the rotation of $\vec{H}$ as it is mainly determined by the angle between $\vec{H}$ and the crystallographic $\hat{c}$ axis whereas the angle between $\vec{H}$ and $\vec{j}$ plays only a minor role, (3) the unusual $T$- and $H$- dependencies of the GMR and the coexistence of large positive and negative MR components where the positive MR component emerges only at low temperatures.

The unusual angular dependence of the GMR defies any explanation using available MR models \cite{pippard1989,son2013,li2012b,yamada1973a,yamada1973b,usami1978a}.
For instance, when a MR originates from the Lorentz force, the magnitude of the $\mr$ is given by the angle between the direction of the charge velocity and the applied $\vec{H}$.
The former is defined by both the symmetry of band structure and the applied $\vec{E}$.
Therefore, the angles that $\vec{H}$ makes with $\vec{j}$ and the specific crystallographic geometry are more or less equally important in shaping the anisotropy of $\mr$ \cite{pippard1989}.
This is clearly inconsistent with the observed small in-plane angular dependence of $\mr (\phi)$ in BMB [Fig.~\ref{fig:anisotropicMR}(d)].
Alternative scenario would be a strong Zeeman-induced Lifshitz transition, i.e., a $\dphyb$ valence band may get spin-split beyond the  small semiconducting gap thus explaining the emergence of electron-like charges and the negative $\mr$ as a function of $\vB$.
Giving the fact that the insulating gap in BMPn's, however being small, is highly anisotropic \cite{zhang2016}, this scenario also seems to be unlikely in the case of BMPn's.

The key to the understanding the GMR in BMPn's thus lays in the unique directional dependence on $\vec{H}$. Namely, MR is predominantly sensitive to the angle between $\vec{H}$ and the crystallographic $\hat{c}$ axis, which is also the easy axis of the G-AFM ordering in BMPn. 
Since BMPn's have very high $T_{\rm N}$ temperatures, all significantly exceeding room temperature, the strongest $H$ of $\SI{50}{\tesla}$ available to our experiments can still be considered as a small perturbation that is unable to induce large changes in the G-AFM order.
An estimate using a mean-field theory shows that in the case of BMBi, the exchange field $H_{\mathrm{exch}}$ 
is around $\SI{250}{\tesla}$, which is more than one order of magnitude larger than $H=\SI{18}{\tesla}$ at which the negative MR is already saturated [Figs.~2(b) and  \ref{fig:BMB_Tdep}(d)].
Due to a very large magnetic anisotropy energy \cite{johnston2011}, the effect of $\vec{H} \perp \hat{c}$ is merely to cant the Mn$^{2+}$ AFM moments by a tiny angle $\epsilon$ from the easy axis as schematically shown in Fig.~\ref{fig:BMB_Tdep}(e).
This is supported by the measurements of magnetic susceptibilities \cite{ogasawara2018,saparov2013a,sangeetha2017,yogeshsingh2009}, in which $\chi_{ab}$ and $\chi_{c}$ show the characteristics expected for an AFM ordering with the $\hat{c}$-axis as the magnetic easy axis \cite{blundell2001magnetism}.
On the other hand, the projections of the AFM spins on the $ab$-plane can easily be rotated with respect to $\vec{H}_{ab}$, giving almost isotropic $\chi_{ab}$.
These magnetic anisotropies perfectly coincide with the symmetry observed in the GMR [Fig.~\ref{fig:BMB_Tdep}(e)], proving that the GMR in BMPn's is directly related to the tilting of the AFM spins from the crystallographic $\hat{c}$ axis.
Although any drastic changes in the G-AFM order, such as large angle spin canting or entering into the spin-flop state, are unlikely to occur for the given experimental field range, the electrical conductivity of BMPn's is still extremely sensitive to the tilting of AFM magnetic sublattices by tiny angles. 

  The sensitivity of the electrical conductivity to the tiny tilting of the AFM moments is interesting from the viewpoint of symmetry breaking.
  The magnetic ground state in BMPn is theoretically accounted for by the $\pt$-symmetric hexadecapolar order \cite{watanabe2017} with the G-AFM leading dipole term.
  A finite $\vec{H}_{ab}$ magnetizes the AFM sublattices in the perpendicular $ab$ plane and thereby removes the $\pt$ symmetry and puts BMPn's into the broken symmetry low resistive state.

  BMPn's are small gap semiconductors, and in general both hole-like and electron-like carriers are important in the electrical conduction.
  With tilting the AFM moments, partial conductivities of both carrier types greatly varies in opposite trends, resulting in a complex $T$ and $H_{ab}$ dependencies of the GMR.  
  A comparison between Figs.~\ref{fig:BMB_Tdep}(a)-(c) and the Hall effects [Figs.~\ref{fig:BMB_Hall}(b)-(c)] indicates a predominance of hole-like carriers in the HT and MT regimes, where a single component negative GMR is observed.
  Only entering LT regime, the positive MR appears in association with the emergent minor electron-like carrier type  
  \footnote{
    The small upward shift of the chemical potential $\mu$ as $T$ decreases [\onlinecite{kittel1980thermal}] can be important in BMPn's due to the narrow band gaps.
    The positive and linear $\rho_{yx}(H_c)$ at high $T$'s [Fig.~\ref{fig:BMB_Hall}(a)] indicates that $\mu$ is located near the top of the valence band so that only the hole-like quasi-particles are excited thermally.
    As $T$ is cooled down, $\mu$ approaches towards the bottom of the conduction band, encouraging the population of the minority electron-like quasi-particles.}.
  Taking into account the sharp contrast in both orbital characteristics and spin-dependent DOS between the $\dphyb$ valence band and the $d$-like conduction band in BMPn's, this observation demonstrates an interesting band-sensitive feature of the GMR.  
  An in-plane $\vec{H}_{ab}$ at the same time hinders the conduction of electron-like carriers occupying the largely spin-down $d$ states and inversely enhances that of the $d\mhyphen p$ holes with almost equal spin up and down populations [Fig.~\ref{fig:xtals_compare}(c)]. 
A small $H_{ab}$ quickly suppresses the conductivity of the high-mobility $d$ electron-like carriers and therefore results in a positive MR.
With increasing $H_{ab}$, the low-mobility $d\mhyphen p$ hole-like carriers, being much larger in number, become more conductive and eventually overwhelm the field-hindered $d$ electron-like carriers, leading to the overall negative GMR in the LT regime.

The high order hexadecapole ground state is microscopically realized via the anisotropic distribution of ``magnetic charges'' via the spin-dependent $d\mhyphen p$ hybridization \cite{watanabe2017}.
Under zero $\vec{H}_{ab}$, the ground state is stabilized with the staggered magnetization parallel to the $\hat{c}$-axis.
On the other hand, by tilting the Mn's spin under a perpendicular $\vec{H}_{ab}$, the distribution of charge and magnetic moment is reset and in turn the $d\mhyphen p$ hybridization will be modified accordingly.
In this picture, a $\vec{H}_{ab}$ magnifies the negative $\mr$ via deforming the top of the valence band and therefore lightening the effective mass of the $\dphyb$ holes.
The observed variations in GMR within the BMPn family can also be understood in terms of $d\mhyphen p$ hybridization and SOC that systematically vary among Pn elements.


  Another plausible mechanism that can be proposed involves an $\vec{H}_{ab}$ enhancement of the carrier mobility by suppression of the spin-mixing transport.
  The PT symmetry renders an almost perfect balanced DOS of majority spin and minority spin states for the low energy hole-like excitations [Fig.~\ref{fig:xtals_compare}(c)].
  With the local magnetization alternatively pointing to opposite $\hat{c}$ and $-\hat{c}$ directions, the special DOS maximizes the mixing between the two spin states in the course of electronic transport either intra- or inter-sublattice
  \footnote{This is in a sharp contrast with those observed in other FM and AFM materials, where a gap usually splits spin majority and spin minority band and carriers tend to stay in the spin-down band \cite{blundell2001magnetism}.}.
  Now, under $H_{ab} \neq 0$ the partially polarization of the AF sublattices (parallel to $\vec{H}_{ab}$) causes a disproportion in the DOS of two spin states.
  Specifically, the states that are spin-down with respect to the direction of the in-plane magnetization can be expected to have large DOS which is shifted up in energy, therefore become more favorable for charge carriers to remain in during the transport.
  Hence the spin-mixing transport, probable to be the direct cause of the low temperature high resistive state, can be conveniently reduced or even turned off by $\vec{H}_{ab}$ and a negative GMR is resulted.
  In this scenario, the magnetotransport property of BMPn's shares the same basic mechanism with the GMR found in artificial AF sublattices \cite{baibich1988,binasch1989}.
  It has been recently shown that by polarizing the AFM spins in a simple Hubbard chain, the GMR will appear in a fashion similar to the context being discussed here \cite{li2018a}.
  We note that here only qualitative proposals for the mechanism of the GMR in BMPn's have been made and a precise description is beyond the scope of this paper.


\section{Summary}
\label{sec:summary}
 We reported the observation of a new type of GMR discovered in the family of the layered BMPn materials.
 The magnitude of the GMR is as large as $\SI{-98}{\percent}$ and its intriguing angular anisotropy under $\vec{H}$ that has not been encountered so far in any other material.
 This GMR has its origin from the interesting tunability of the highly anisotropic spin-dependent $d\mhyphen p$ hybridization and spin dependent scatterings.
 Most remarkably, via the spin-charge coupling the dynamics of charge carriers was able to be greatly altered when the AFM magnetization $\vec{M}$ is tilted with $\vec{H}$ even for a very small angle.
The novel MR mechanism described here is interesting by recalling the recent reports of the dynamical ME effect, in which electrical switching of magnetic ordering was experimentally demonstrated in itinerant AFM \cite{wadley2016,olejnik2017}.
However, its inverse phenomenon, as to how an electrical conduction quantity is modified when  the ground state of materials is magnetically perturbed, has not yet been elucidated experimentally and theoretically.
The described novel GMR may shed light on the inverse ME effect of AFM materials with two magnetic sublattices preserving the PT symmetry in the itinerant regime.

\section*{Methods}
\label{sec:experiment}
We synthesized single crystals of four compounds in the BMPn family via flux method.
For BMSb and BMBi, Sb and Bi metallic fluxes were used, respectively \cite{saparov2013}.
BMAs single crystals were growth from a conventional MnAs self-flux.
BMP crystals were growth from Ba$_x$P$_y$ flux \cite{nakajima2012}.
The single crystals of all three compounds exhibit shiny mirror $ab$ plane with clear facets.
It is important to note that whereas BMA is stable, BMBi and BMSb are sensitive to air; the colors of those crystals gradually change after around several hours exposing.

In the measurements of in-plane electrical transport properties, the crystals were cleaved to expose fresh $(001)$ surfaces and cut into rectangular shapes.
Four electrodes were made on the $ab$ plane and along the $[100]$ direction by silver paste for resistivity measurements.
Two additional Hall electrodes were made on the two opposite sides of the samples.
The two current electrodes at the two opposite ends of the sample were made so that the silver paste covered all the edge sides to ensure a good current flow \cite{pippard1989}.
All the sample preparation typically took less than two hours.

Directional magnetotransport measurements were carried out under magnetic field strength ($H$) smaller than $\SI{9}{\tesla}$ using a Quantum Design PPMS equipped with a rotator option.
The experiments under higher magnetic fields were carried out at High Field Laboratory at Institute for Materials Research, Tohoku University with the help of a $\SI{18}{\tesla}$ superconducting magnet.
The verification of the saturation behavior in $\mr$'s under extreme $H$ was achieved via measurements employing a $\SI{55}{\tesla}$ pulse magnet at Center for Advanced High Magnetic Field Science, Osaka University.
In the case of BMAs, the large positive $\mr$ of the sample made the pulse-$H$ data under $\mu_0H < \SI{30}{\tesla}$ very difficult; the value of $\mr$ shown in Fig.~\ref{fig:BMPn_MRs}(b) was calculated by comparing the values of resistivity under $H = 0$ and $\mu_0 H \geq \SI{30}{\tesla}$.
In order to describe the complex dependencies on both direction and magnitude of $\vB$, it is convenient to define the isothermal $\mr$ measured at a specific $T$ as,
\begin{align*}
  \label{eq:MRdef}
  \mr(T, H, \theta^{\prime}, \phi) = \frac{\rho_{xx}(T, H, \theta^{\prime}, \phi)}{\rho_{0}(T)} - 1\,,
\end{align*}
where $H$ is the magnitude of $\vec{H}$ and $\rho_0(T)$ the value of $\rho$ at $T$ under $H = 0$.
As illustrated in figures \ref{fig:anisotropicMR}(a) and (c), $\theta^{\prime}$ and $\phi$ are the angles between $\vec{H}$ and the $[100]$ ($\hat{a}$-) direction in the $ac$ and $ab$ plane, respectively.
The notations $\vB_{ab}$ denote a $\vB$ that is parallel to the $ab$ plane.
The notations $H_{ab}$, $H_{a}$, and $H_c$ denote the magnitude of a $\vB$ (or the component of $\vB$) that is parallel to the $ab$ plane, $\hat{a}$-, and $\hat{c}$-axis, respectively.
In the transverse Hall measurements, $\vec{H}$ was always aligned along the $\hat{c}$ axis, and $\rho_{yx}$ was measured as a function of the Hall field strength $H_c$.

\section*{Acknowledgements}
We thank K. Ogushi, T. Aoyama, K. Igarashi, H. Watanabe, Y. Yanase, and T. Arima for fruitful discussions. We also thank S. Kimura and S. Awaji for excellent support in the measurements.
This work was supported by a Grant-in-Aid for Scientific Research on Innovative Areas ``J-Physics'' (Grant No.18H04304), and by JSPS KAKENHI (Grants No. 18K13489, No. 18H03883, No. 17H045326, and No. 18H03858).
Pulsed field measurements were carried out at the Center for Advanced High Magnetic Field Science
in Osaka University under the Visiting Researcher's Program of the Institute for Solid State
Physics, the University of Tokyo.
This research was partly made under the financial support by the bilateral country research program of JSPS between AIMR, Tohoku University and Jozef Stefane Institute, Slovenia.
This work was also supported by World Premier International Research Center Initiative (WPI), MEXT, Japan. DA acknowledges the financial support of the Slovenian Research Agency through BI-JP/17-19-004 and J1-9145 grants.


\bibliography{BaMn2Pn2_related}

\end{document}